\documentclass[%
 reprint,
 amsmath,amssymb,
 aps, 
 prl, 
 superscriptaddress, 
]{revtex4-2}

\usepackage{siunitx}
\DeclareSIUnit\angstrom{\text {Å}}
\usepackage{graphicx}
\usepackage{dcolumn}
\usepackage{comment}
\usepackage{float} 
\usepackage{xcolor}
\usepackage{ulem}
\usepackage{bm}
\usepackage{hyperref}

\usepackage{soul}
\usepackage{pdfpages}
\makeatletter
\AtBeginDocument{\let\LS@rot\@undefined}
\makeatother

\begin{document}

\preprint{APS/123-QED}

\title{Emergent Symmetry and Valley Chern Insulator in Twisted Double-Bilayer Graphene}

\author{Yimeng Wang}
\affiliation{
  Microelectronics Research Center, Department of Electrical and Computer Engineering, The University of Texas at Austin, Austin, TX 78758, USA
}

\author{G. William Burg}
\affiliation{
  Microelectronics Research Center, Department of Electrical and Computer Engineering, The University of Texas at Austin, Austin, TX 78758, USA
}

\author{Biao Lian}
\affiliation{
Princeton Center for Theoretical Science, Princeton University, Princeton, New Jersey 08544, USA
}

\author{Kenji Watanabe}
\affiliation{
  Research Center for Functional Materials, National Institute of Materials Science, 1-1 Namiki Tsukuba, Ibaraki 305-0044, Japan
}
\author{Takashi Taniguchi}
\affiliation{
  International Center for Materials Nanoarchitectonics, National Institute of Materials Science, 1-1 Namiki Tsukuba, Ibaraki 305-0044, Japan
}
\author{B. Andrei Bernevig}
\affiliation{
Department of Physics, Princeton University, Princeton, New Jersey 08544, USA
}
\author{Emanuel Tutuc}
\email[Corresponding Author: ]{etutuc@mail.utexas.edu}
\affiliation{
  Microelectronics Research Center, Department of Electrical and Computer Engineering, The University of Texas at Austin, Austin, TX 78758, USA
}

\date{\today}

\begin{abstract}
Theoretical calculations show that twisted double bilayer graphene (TDBG) under a transverse electric field develops a valley Chern number $2$ at charge neutrality. Using thermodynamic and thermal activation measurements we report the experimental observation of a universal closing of the charge neutrality gap in the Hofstadter spectrum of TDBG at $1/2$ magnetic flux per unit cell, in agreement with theoretical predictions for a valley Chern number $2$ gap. Our theoretical analysis of the experimental data shows that the interaction energy, while larger than the flat-band bandwidth in TDBG near $1^\circ$ does not alter the emergent valley symmetry or the single-particle band topology. 
\end{abstract}

\maketitle

Moir\'e patterns of two-dimensional (2D) materials provide a highly tunable platform for exploring correlated electronic states in electronic bands with flat dispersion. 
Celebrated examples include the twisted bilayer graphene (TBG) at the magic angle $\theta \simeq 1.1^\circ$\cite{bistritzer2011,cao2018,cao_correlated_2018,yankowitz_tuning_2019,lu_superconductors_2019}, twisted double-bilayer graphene  \cite{liu_spin-polarized_2019,burg_correlated_2019,shen_correlated_2020,cao_tunable_2020}, ABC trilayer graphene on hexagonal boron-nitride (hBN) \cite{chen_evidence_2019,chen_signatures_2019}, twisted trilayer graphene \cite{park2021c,hao2021a,cao2021,kim2022}, as well as fractional Chern insulators in twisted bilayer MoTe$_2$ \cite{zeng2023,cai2023,xu2023} and pentalayer rhombohedral graphene on hBN \cite{lu2024}, all of which exhibit superconducting or correlated insulator phases. 
Theoretical calculations show that a new emergent symmetry - which decouples two originally strongly intertwined valleys \cite{bistritzer2011} - appears, and that, per each of the two valley emergent symmetry sectors, the electron bands of many moir\'e systems are topologically nontrivial. For example, 
TDBG and ABC trilayer graphene on hBN carry non-zero emergent-valley Chern numbers $2$ and $3$, respectively, under a transverse electric field  \cite{chebrolu2019,koshino2019,zhangy2019,liuj2019,chittari2019,lee2019}. Unlike correlated Chern insulators induced by interactions in the flat moir\'e bands, the valley Chern insulators in these systems stem from the nontrivial band topology protected by time-reversal invariance.

Experimental signatures of nontrivial band topology include transport experiments of the quantum spin Hall effect in 2D time-reversal invariant systems \cite{konig2007}, the quantum anomalous Hall effect of Chern insulators in magnetic topological insulator thin films \cite{chang2013}, and the angle-resolved photoemission spectroscopy and scanning tunneling microscope studies of 3D time-reversal invariant topological insulators \cite{xia2009,cheny2009,zhangt2009}. 
In TDBG under transverse electric field the gap at charge neutrality carries a topological valley Chern number $C_V=2$, protected by an emergent valley U(1) symmetry \cite{chebrolu2019,koshino2019,zhangy2019,liuj2019,lee2019}. This emergent symmetry may be broken on the edge, leading to back-scattering  between the counter-propagating edge states from different valleys. Although interactions may drive moir\'e systems into a correlated Chern insulator with protected edge states \cite{sharpe_emergent_2019,serlin2020,chen_tunable_2020,wang2022,cai2023,xu2023,lu2024}, such state is not necessarily related to the non-interacting band topology protected by the emergent valley symmetry. A bulk measurement of the band topological character per emergent symmetry sector (valley) is therefore necessary. 

Theoretical studies suggest that the nontrivial band topology manifests itself in the Hofstadter butterfly, and can be observed through bulk magneto-transport \cite{lian2018,wu2021a,crosse2020,lian2020,arbeitman2020}. In particular, a non-interacting band gap that carries a topological valley Chern number $C_V>0$ is predicted to close at or before magnetic flux per unit cell $\Phi=(1/C_V)\Phi_0$ ($\Phi_0=h/e$ is the flux quanta). Here, we present a combined experimental and theoretical study of magneto-transport in TDBG at $\theta=0.97^\circ$, $1.01^\circ$ and $1.33^\circ$. As a function of perpendicular magnetic field ($B$) the experimental data show an unexpected closing of the gap at charge neutrality when the flux per moir\'e unit cell $\Phi\approx(1/2)\Phi_0$ or smaller, in contrast with the tenets of the quantum Hall effect, which dictate that gaps increase as a function of $B$. This observation is consistent with the valley Chern number $C_V=2$ of the valley-filtered flat bands, which supports both the band topology and the emergent symmetry that protects it. Furthermore, by theoretically analyzing the effect of interactions, we show that our experiment implies an interaction energy scale larger than the bandwidth of the flat bands in TDBG at $\theta=1.01^\circ$, and that the interaction does not spoil the single-particle band topology and the emergent valley symmetry at charge neutrality.

\begin{figure}[!htbp]
\begin{center}
\includegraphics[width=3.4in]{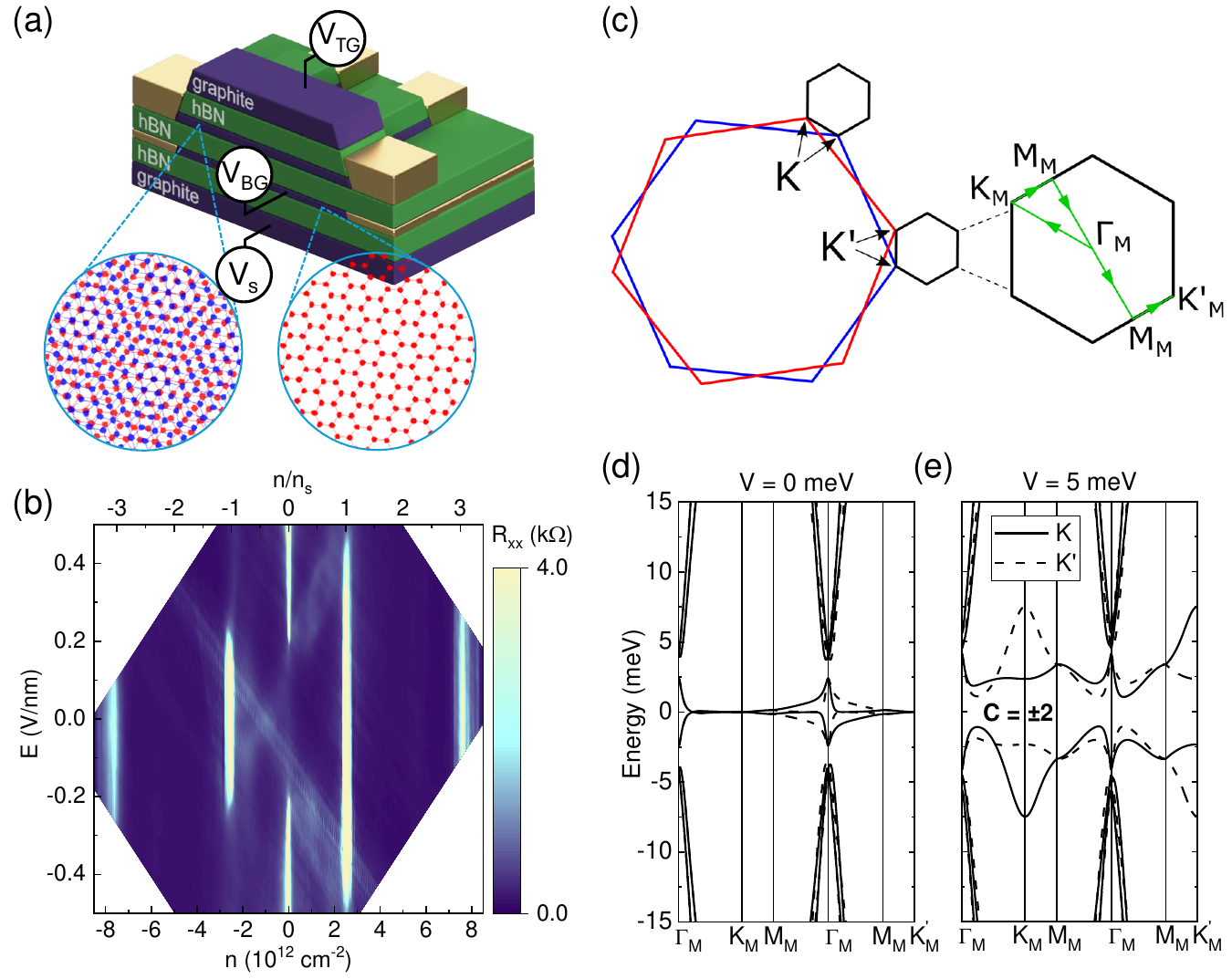}
\end{center}
\caption{(a) TDBG sample schematic with graphite top gate and graphene bottom gate. For chemical potential measurements the bottom gate is made of monolayer graphene and has multiple contacts. A substrate gate below the bottom gate is used to tune the $E$-field at which the chemical potential is measured. (b) Contour plot of $R_\mathrm{xx}$ vs. $n$ and $E$ measured in the TDBG sample with $\theta=1.01^\circ$ at $T=\SI{1.5}{K}$. The top axis shows $n$ in units of $n_\mathrm{s}$. (c) Decoupled moir\'e Brillouin zones arise at the $K$ and $K'$ points of the graphene bilayers. The enlarged view of the moir\'e Brillouin zone shows the high symmetry points and the path taken in momentum space for the (d-e) calculations. (d-e) Calculated moir\'e band structures at (d) $V=\SI{0}{meV}$ and (e) $V=\SI{5}{meV}$. Solid (dashed) lines represent the moir\'e bands in the $K$ ($K'$) valleys of the graphene bilayers. At finite $V$, the gap at charge neutrality has Chern numbers $+2$ and $-2$ in the $K$ and $K'$ valleys, respectively.}
\label{fig1}
\end{figure}

Figure 1(a) shows a schematic of the TDBG sample, with hBN dielectrics and three gates -- top, bottom, and substrate gates. The top and the bottom gates are used to independently control the carrier density ($n$) and transverse electric field ($E$) in the TDBG. For the $\theta=0.97^\circ$ sample the bottom gate, which is made of monolayer graphene and contacted with several electrodes, also acts as a resistively detected Kelvin probe to the TDBG chemical potential. The substrate gate is added to the dual gated geometry to tune the transverse $E$-field at which the TDBG chemical potential is measured \cite{wang2022}.

Figure 1(b) shows the longitudinal resistance ($R_\mathrm{xx}$) as a function of $n$ and $E$ in a TDBG sample with $\theta = 1.01^\circ$. The twist angle is defined such that in the limit $\theta = 0^\circ$ the TDBG approaches AB-AB stacking [Fig. 1(c)].  $R_\mathrm{xx}$ maxima along lines of constant $n$ indicate energy gaps in the moir\'e band structure. The data show $E$-dependent, single particle gaps at charge neutrality ($n=0$) and at fixed densities $\pm n_\mathrm{s}$ and $\pm 3n_\mathrm{s}$. Here, $n_\mathrm{s}$ is the carrier density required to fill a moir\'e unit cell with 4-fold spin-valley degeneracy, and is related to the twist angle by 
\[n_\mathrm{s} = 4\frac{2}{\sqrt{3}}\left[ \frac{\sin(\frac{\theta}{2})}{(a/2)} \right]^2\mathrm{;}\] $a = \SI{2.46}{\angstrom}$ is the graphene lattice constant. At charge neutrality, the band gap is initially small and increases with $E$, while the gaps at $\pm n_\mathrm{s}$ decrease and eventually close at finite $E$. In addition, there are emerging $R_\mathrm{xx}$ maxima at $n=+\frac{1}{2}n_\mathrm{s}$ within small windows of $E$ around $\pm \SI{0.3}{V/nm}$, corresponding to developing correlated insulators at half-filling of the first conduction moir\'e band. Correlated insulators appear at half and quarter moir\'e band filling in TDBG flat bands as a result of strong electron-electron interaction \cite{liu_spin-polarized_2019,burg_correlated_2019,shen_correlated_2020,cao_tunable_2020,he_tunable_2020}.

For small $\theta$ the TDBG moir\'e pattern reciprocal vector is much smaller than the intervalley momentum difference, which leads to distinct moir\'e Brillouin zones at the $K$ and $K'$ valleys of the parent graphene bilayers that are exponentially (in $1/\theta$) decoupled from one another [Fig. 1(c)]. This decoupling gives the moir\'e bands a valley degree of freedom, and an emergent valley U(1) rotation symmetry. Figures 1(d)-1(e) show $\theta = 1.01^\circ$ band structure calculations for zero and finite on-site energy difference ($V$) between neighboring graphene layers, respectively, and demonstrate the valley dependence of the bands. The $V$ values are proportional to the applied transverse electric field $E$ \cite{suppmat}. Furthermore, the calculated $V$-dependence is consistent with experimental observations, showing that the gap at charge neutrality opens with increasing $E$, and the gaps between the first and second moir\'e bands reduce or close. 

\begin{figure*}[t]
\begin{center}
\includegraphics[width=6in]{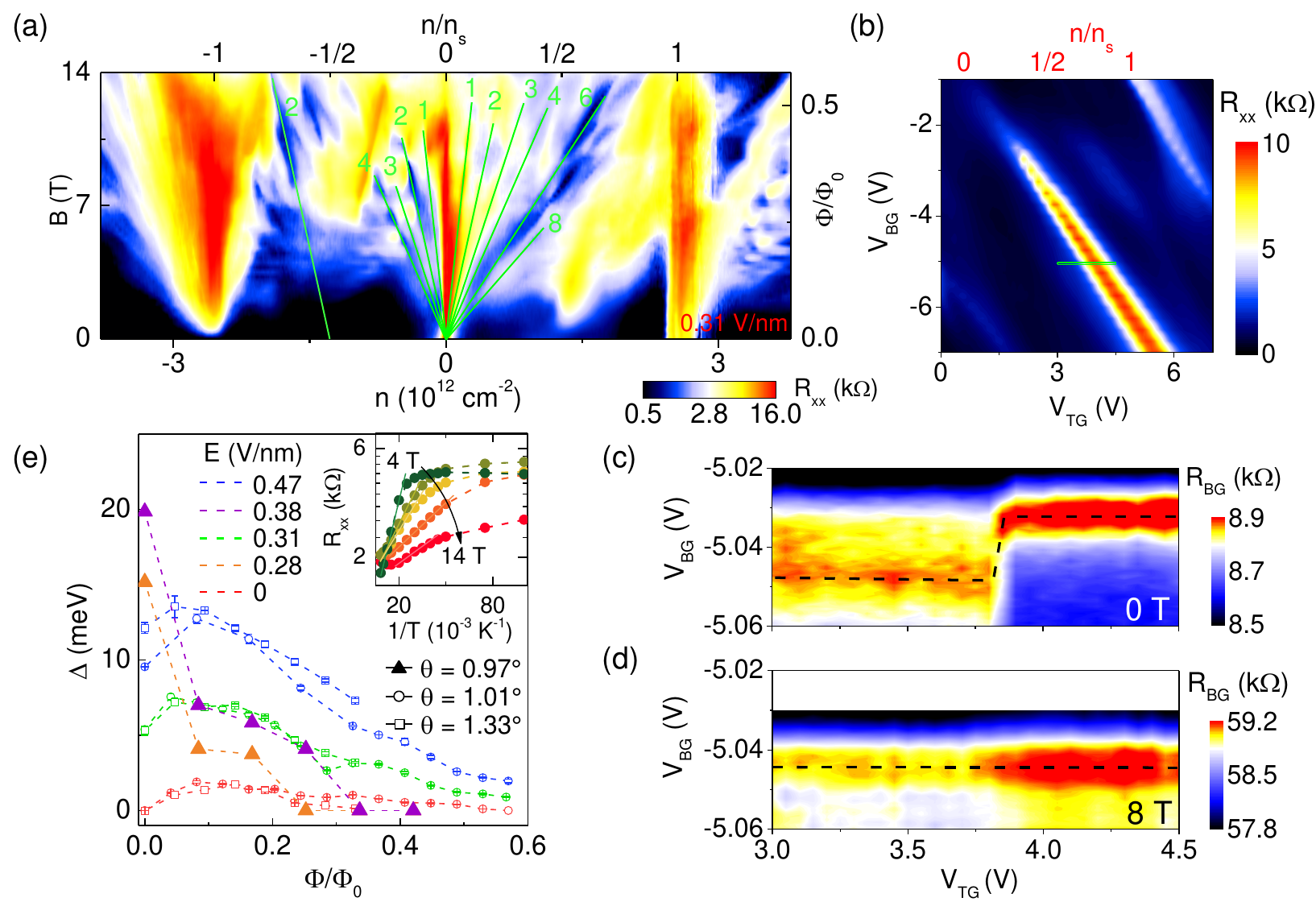}
\end{center}
\caption{(a) $R_\mathrm{xx}$ vs. $n$ and $B$ at $E=\SI{0.31}{V/nm}$ for TDBG with $\theta = 1.01^\circ$. The top axis shows $n$ in units of $n_\mathrm{s}$, and the right axis shows $\Phi/\Phi_\mathrm{0}$. The Landau level filling factor for well developed quantum Hall states are labeled. (b) $R_\mathrm{xx}$ vs. $V_\mathrm{TG}$ and $V_\mathrm{BG}$ for the $\theta = 0.97^\circ$ TDBG. The top axis shows $n/n_\mathrm{s}$ for the diagonal resistance maxima. (c-d) $R_\mathrm{BG}$ vs. $V_\mathrm{TG}$ and $V_\mathrm{BG}$ measured at (c) $B = \SI{0}{T}$ and (d) $\SI{8}{T}$. Black dashed lines mark charge neutrality loci of the bottom gate graphene. Data are taken at $T=$ \SI{1.5}{K} in (a-d). (e) $\Delta$ vs. $\Phi/\Phi_\mathrm{0}$ at different $E$-fields for three TDBG samples. The triangles represent thermodynamic measurements in the $\theta = 0.97^\circ$ TDBG; the circles and squares represent thermal activation measurements in the $\theta = 1.01^\circ$ and $1.33^\circ$ TDBGs. Inset shows Arrhenius plot of $R_\mathrm{xx}$ vs. $1/T$ at charge neutrality for $E=\SI{0.47}{V/nm}$, and at different $B$-fields for the $\theta = 1.01^\circ$ TDBG.}
\label{fig2}
\end{figure*}

At finite $V$, the gap that opens at charge neutrality is predicted to be topologically nontrivial \cite{chebrolu2019,koshino2019,zhangy2019,liuj2019,lee2019}, with a valley-dependent Chern number of $C_K=+2$ or $C_{K'}=-2$ [Fig. 1(e)], which originates from the opening of two double Dirac cones of the same helicity in each valley.  The gap thus carries a valley Chern number $C_V=(C_K-C_K')/2=2$  per spin, protected by the emergent valley U(1) symmetry, a bulk symmetry that can be broken by edge roughness. Instead, we explore here bulk signatures of the nonzero valley Chern number by examining the magneto-transport of TDBG in perpendicular magnetic fields, with an emphasis on the gap at charge neutrality. Figure 2(a) shows $R_\mathrm{xx}$ as a function of $n$ and $B$ for $E = \SI{0.31}{V/nm}$ in $\theta=1.01^\circ$ TDBG. The combination of perpendicular magnetic fields and the spatial modulation of the moir\'e pattern leads to equally spaced Landau levels in the energy spectrum, and quantum Hall states (QHSs) when $n$ and $\Phi$ satisfy the Streda formula \cite{streda1982, bistritzer2011b}: 
\[\frac{n}{n_\mathrm{s}}=\frac{\nu}{4} \frac{\Phi}{\Phi_\mathrm{0}}+\sigma\mathrm{,}\] 
where $\Phi = BA_M$, $A_M=4/n_s$ is the moir\'e unit cell area, and $\nu$ and $\sigma$ are Landau level and moir\'e band filling factors, respectively. The best developed QHSs are labeled by their corresponding $\nu$ value in Fig. 2(a). An advantage of small twist angle TDBG is the large unit cell area, which makes high magnetic flux values relatively easy to access experimentally. As shown on the right axes of Fig. 2(a), $\Phi/\Phi_\mathrm{0}$ exceeds 1/2 at $B=\SI{14}{T}$, values at which the predicted topological signatures should manifest.

Focusing now on the $\Phi/\Phi_\mathrm{0}$ dependence of the gap at charge neutrality, Fig. 2(a) shows that $R_\mathrm{xx}$ at charge neutrality is initially large, and generally independent of $B$ for low and moderate fields, but then decreases as $\Phi/\Phi_\mathrm{0}$ approaches 1/2. While resistance is not necessarily a direct measure of an energy gap, especially when topological edge states may be present, these findings suggest that the gap at charge neutrality decreases at high magnetic flux in the vicinity of $\Phi/\Phi_\mathrm{0}=1/2$. To substantiate this finding, a bulk measurement is needed. To that end, we measure the gap ($\Delta$) at charge neutrality as a function of $\Phi/\Phi_\mathrm{0}$ using a combination of thermodynamic measurements and thermal activation. Figure 2(b)-2(d) show the thermodynamic measurements in the $\theta = 0.97^\circ$ TDBG. In Fig. 2(b), $R_\mathrm{xx}$ as a function of $V_\mathrm{TG}$ and $V_\mathrm{BG}$ is shown, where the resistance maxima along the diagonal represent the TDBG charge neutrality. Figure 2(c) shows the resistance of the graphene bottom gate $R_\mathrm{BG}$ as a function of $V_\mathrm{TG}$ and $V_\mathrm{BG}$ measured at $E = \SI{0.38}{V/nm}$ and $B = \SI{0}{T}$. The $R_\mathrm{BG}$ is measured around the TDBG charge neutrality in the range marked by the green rectangle in Fig. 2(b), with the substrate gate at a fixed bias to shift the neutrality point of the bottom gate graphene to the target $E$-field. As detailed in Ref. \cite{wang2022}, by tracing the charge neutrality of the bottom gate graphene, the chemical potential of the TDBG can be measured directly, revealing the bulk properties at charge neutrality. Specifically, the chemical potential of the TDBG can be calculated with $\mu = eV_\mathrm{BG}(1+C_\mathrm{s}/C_\mathrm{BG})-eV_\mathrm{s}C_\mathrm{s}/C_\mathrm{BG}$. Here, $C_\mathrm{BG}$ and $C_\mathrm{s}$ are the capacitance per unit area of the bottom and substrate gate dielectric, respectively, and $V_\mathrm{s}$ is the substrate gate bias. Accordingly, the change of $\mu$ is proportional to the change of $V_\mathrm{BG}$ when tracing the bottom gate charge neutrality plotted in Fig. 2(c), and the step the charge neutrality loci take at $V_\mathrm{BG} \approx \SI{3.8}{V}$ reflects the TDBG thermodynamic gap at its charge neutrality, which is $\Delta =\SI{20}{meV}$. The data for the same measurement at $B = \SI{8}{T}$ are shown in Fig. 2(d), where no gap is observed. The values of $\Delta$ vs. $\Phi/\Phi_\mathrm{0}$ measured in the $\theta = 0.97^\circ$ TDBG at fixed $E$-field values are shown in Fig. 2(e) (triangle symbols). Meanwhile, we perform thermal activation measurements on the $\theta = 1.01^\circ$ and $1.33^\circ$ TDBGs [Fig. 2(e) inset], and summarize the thermodynamic and activation measurement results of $\Delta$ vs. $\Phi/\Phi_\mathrm{0}$ in Fig. 2(e). Although differences exist between datasets, in both types of measurements the charge neutrality gap shows a closing behavior near or before $\Phi/\Phi_\mathrm{0} = 1/2$, which suggests a common origin of the gap closing. In the $\theta=1.33^\circ$ TDBG, $\Delta$ shows a similar decreasing trend, but is only measurable up to $\Phi/\Phi_\mathrm{0} \simeq 0.3$ due to the larger twist angle.

\begin{figure}[!htbp]
\begin{center}
\includegraphics[width=3.3in]{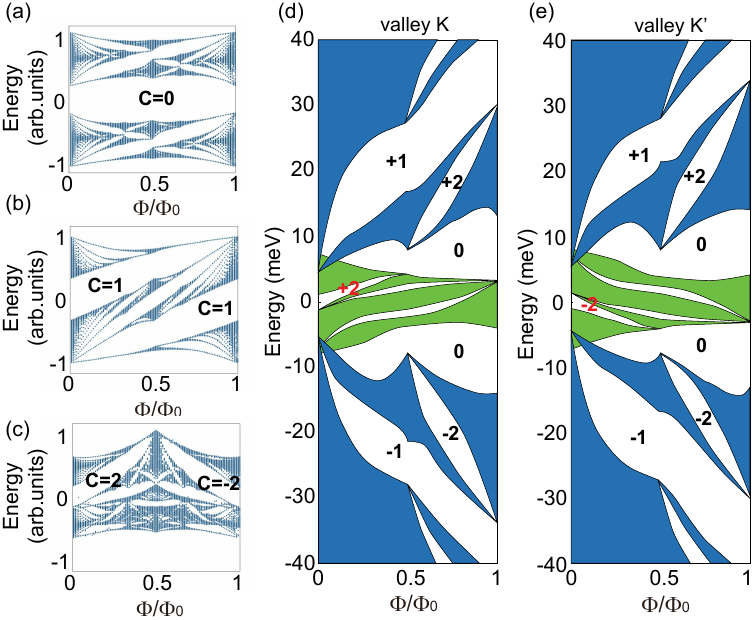}
\end{center}
\caption{(a-c) Hofstadter butterflies of 2-band Chern insulator tight-binding models with Chern numbers $0,1$ and $2$, respectively. A gap of Chern number $C$ is topologically protected to close at $\Phi/\Phi_0=1/C$. (d-e) The Hofstadter spectra of TDBG with $\theta=1.01^\circ$ and $V=5$ meV at valleys $K$ and $K'$, respectively. The spectra include the lowest two moir\'e bands (green) and the higher moir\'e bands (blue). The integers label the Chern numbers of the dominant Hofstadter gaps.}
\label{fig3}
\end{figure}

The observations in Fig. 2 are surprising. For typical band-structures energy gaps are expected to open with increasing $B$.
In magic angle TBG, a similar system with flat bands and strong interactions, compressibility measurements reveal the charge neutrality gap increases with $B$ \cite{yu2022}. Consequently, as we explain below, the observed gap closing near $\Phi/\Phi_\mathrm{0}=1/2$ is compelling evidence of the nontrivial topology of charge neutrality gap in TDBG, and of the emergent symmetry that protects the valley Chern number. It is important to distinguish between the valley Chern insulator in TDBG, and the correlated Chern insulators in TBG that break time reversal symmetry \cite{nuckolls2020b,saito2021, xie2021a}. The fillings of the latter disperse in a perpendicular magnetic field according to the Streda formula \cite{streda1982, bistritzer2011b}, similar to a quantum Hall state. The valley Chern gap in TDBG does not break time reversal symmetry, and is not expected to disperse according to the Streda formula.

In the non-interacting band theory, the Hofstadter spectrum of a trivial band is usually bounded within its bandwidth, leaving the gap between topologically trivial bands open at any $\Phi/\Phi_0$ as shown in Fig. 3(a) \cite{suppmat}. In contrast, a topological gap will generically close at a certain magnetic flux, connecting the Hofstadter spectra above and below the gap. The simplest example is a Chern number $C$ gap between two Chern bands of Chern numbers $\pm C$, which is topologically enforced to close at $\Phi/\Phi_0=1/|C|$ \cite{suppmat,lian2018,lian2020,arbeitman2020}. Two examples of Chern number $C=1$ and $C=2$ gaps are shown in Fig. 3(b) and 3(c), respectively \cite{suppmat}.

In TDBG with nonzero $V$, each valley has a Chern number $\pm2$ gap at charge neutrality, which is forced to close at $\Phi/\Phi_0=1/2$. Fig. 3(d)-3(e) show the Hofstadter butterflies at valley $K$ and $K'$ calculated \cite{lian2018,lian2020} using the non-interacting TDBG continuum model at $\theta=1.01^\circ$ and $V=\SI{5}{meV}$. The spectra of the lowest two moir\'e bands are shown in green. As expected, at $\Phi/\Phi_0=1/2$, the Chern number $C_K=2$ ($C_{K'}=-2$) charge neutrality gap of valley $K$ ($K'$) closes at the top (bottom) of the lowest two moir\'e bands. Figure 4(a) shows the single valley gap with respect to $\Phi/\Phi_0$ at different $V$. This strongly suggests that the experimentally observed charge neutrality gap closing at $\Phi/\Phi_0=1/2$ in TDBG is due to the valley-dependent Chern number $C_K=-C_{K'}=2$. 

However, the midgap energies of the $C_K=2$ and $C_{K'}=-2$ Chern gaps at valleys $K$ and $K'$ disperse oppositely with $\Phi$, which differ by $\sim10$meV at $\Phi/\Phi_0=1/2$ [Fig. 3(d)-3(e)]. Furthermore, the Zeeman spin splitting is $\sim1$meV at $\Phi/\Phi_0=1/2$. Therefore, the total single-particle valley Chern gap of all spins and valleys at charge neutrality would close before $\Phi/\Phi_0=1/2$ due to energy band overlaps, as illustrated in Fig. 4(b). The calculated total valley Chern gap $\Delta$ of all spins and valleys for TDBG at $\theta=1.01^\circ$ and different $V$ closes no later than $\Phi/\Phi_0\approx 0.2$ [Fig. 4(c)], and shows a gap reopening around $0.2<\Phi/\Phi_0< 0.4$ due to other, further, Landau level gaps, different from our experimental observation. This discrepancy can be resolved by considering electron-electron interactions, which we treat here in a mean-field theory. Denote the band bottom and top of the first conduction (valence) moir\'e band at $\Phi=0$ as $M_1$ ($-M_0$) and $M_0$ ($-M_1$), respectively, as shown in Fig. 4(b) ($M_0> M_1> 0$). For simplification, we assume the Chern gap of valley $\eta$ ($+1$ for $K$ and $-1$ for $K'$) decreases linearly and closes at energy $\eta M_0$ at $\Phi/\Phi_0=1/2$. The number of conduction (valence) states per area above (below) the gap at valley $\eta$ and spin $s$ is given by $N_{m,\eta}(\Phi)=(1-2m\eta\Phi/\Phi_0)/A_M$ ($m=\pm1$ for conduction and valence bands, respectively).

\begin{figure}[!htbp]
\begin{center}
\includegraphics[width=3.4in]{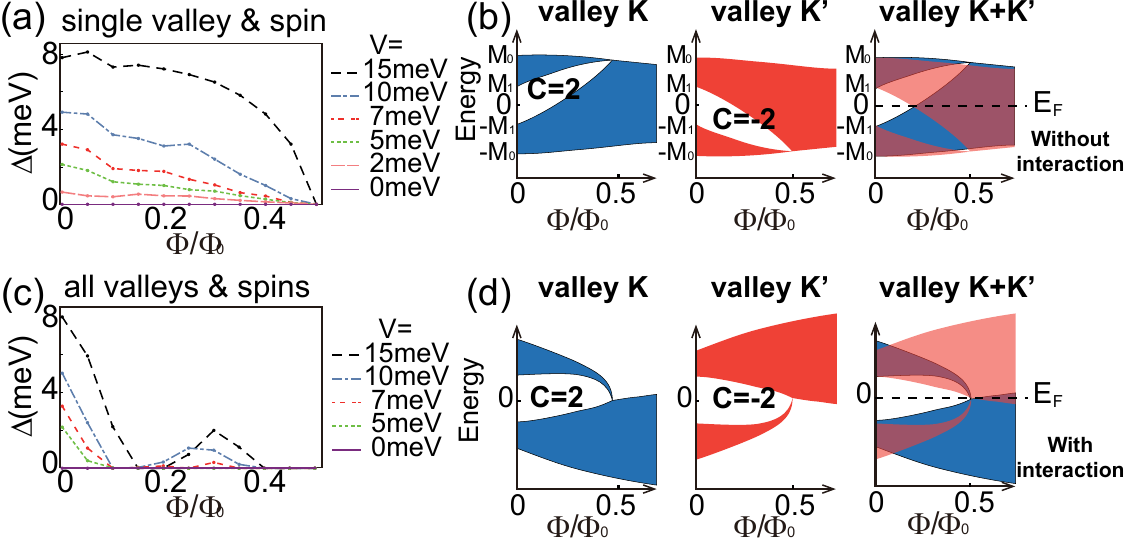}
\end{center}
\caption{(a) The $C_\eta=\pm2$ Chern gap of a single valley and spin vs. $\Phi/\Phi_0$. (b) Schematics of the non-interacting Hofstadter spectra at valley $K$ and $K'$. The $C_\eta=\pm2$ Chern gaps of the two valleys ($\eta=\pm1$ for valleys $K$ and $K'$, respectively) split in energy at finite magnetic field, leading to a total charge neutrality gap closing before $\Phi/\Phi_0=1/2$. (c) The total non-interacting charge neutrality gap of all spins and valleys of TDBG vs. $\Phi/\Phi_0$, where spin g-factor $2$ is used. (d) Illustration of the effective Hofstadter spectra under interaction by mean field theory, where the $C_\eta=\pm2$ Chern gaps of valleys $K$ and $K'$ are enlarged and shifted towards each other, leading to a total charge neutrality gap closing around $\Phi/\Phi_0=0.5$.}
\label{fig4}
\end{figure}

We consider a local interaction particle-hole symmetric about charge neutrality: 
\[
\mathcal{H}_I=\frac{U_0A_M}{2}\sum_{m,m',\eta,\eta',s,s'}\gamma_{m \eta s}^{m'\eta's'}(\Phi)\widetilde{n}_{m,\eta,s}\widetilde{n}_{m',\eta',s'},
\]
where $U_0>0$ is the short-range interaction energy,  $\widetilde{n}_{m,\eta,s}=n_{m,\eta,s}-\frac{N_{m,\eta}(\Phi)}{2}$ is the electron density of indices $\{m,\eta,s\}$  relative to half-filling, with $0\le n_{m,\eta,s}\le N_{m,\eta}(\Phi)$, and $\gamma_{m \eta s}^{m'\eta's'}(\Phi)=1-\delta_{m,m'}\delta_{\eta,\eta'}\delta_{s,s'}-\delta_{m,-m'}\delta_{\eta,\eta'}\delta_{s,s'}\zeta(\Phi)$ accounts for reduced interaction within the same valley and spin due to Pauli exclusion \cite{suppmat}. We assume each band $\{m,\eta,s\}$ has a uniform density of states. This allows us to calculate the total mean-field energy at the filling of charge neutrality, and derive the ground-state charge neutrality gap $\Delta(\Phi)$ with respect to $\Phi$ \cite{suppmat}. We find two interaction effects: first, the charge neutrality gap at $\Phi=0$ increases from the single-particle gap $2M_1$ to $\Delta(0)=2M_1+U_0$. Secondly, the charge-neutrality valley Chern gap $\Delta(\Phi)$ closes at $\Phi/\Phi_0=\frac{2M_1+U_0}{4M_0+4M_1-2U_0}<1/2$ for weak interactions $0<U_0<M_0$ (but is larger than non-interacting cases), while the gap closes at $\Phi/\Phi_0\approx1/2$ for strong interactions $U_0\ge M_0$, analogous to the single valley gap in Fig. 4(a). Heuristically, this is because the interaction increases the energy cost for a valence electron to jump into a conduction band, stabilizing the charge neutrality gap. In the mean field picture, the bands of the two valleys are effectively deformed by interaction as shown in Fig. 4(d), with their Chern gaps enlarged and shifted towards each other. Such a simplified mean field picture is further verified by our numerical Hartree-Fock calculation of the Hofstadter spectrum of a valley Chern number $2$ tight-binding model with on-site Hubbard interactions \cite{suppmat}.

Our experimental observations reveal the TDBG charge neutrality gap under a transverse $E$-field shows an unusual closing in the presence of a perpendicular magnetic field, consistent with a valley Chern insulator with $C_V=2$.  
Moreover, the interaction energy $U_0$, although larger than the flat band widths in TDBG near $\theta=1^\circ$, preserves the emergent valley U(1) symmetry and the valley Chern number. 
Our study provides a novel method to detect the band topology of 2D moir\'es and other superlattice systems by measurement of their bulk energy spectra.
 
\begin{acknowledgments}
The work at The University of Texas was supported by the National Science Foundation Grants MRSEC DMR-2308817 and EECS-2122476, Army Research Office under Grant No. W911NF-22-1-0160, and the Welch Foundation grant F-2169-20230405. Work was partly done at the Texas Nanofabrication Facility supported by NSF Grant No. NNCI-1542159. B.A.B. was supported by the Department of Energy Grant No. de-sc0016239, the Schmidt Fund for Innovative Research, Simons Investigator Grant No. 404513 the Packard Foundation. B.L. is supported by the National Science Foundation through Princeton University’s Materials Research Science and Engineering Center DMR-2011750, and the National Science Foundation under award DMR-2141966. K.W. and T. T. acknowledge support from the Elemental Strategy Initiative conducted by the MEXT, Japan (Grant No. JPMXP0112101001) and JSPS KAKENHI (Grants No. JP19H05790 and No. JP20H00354).
\end{acknowledgments}


\bibliographystyle{apsrev4-2}
%


\clearpage
\newpage
\includepdf[pages={1}]{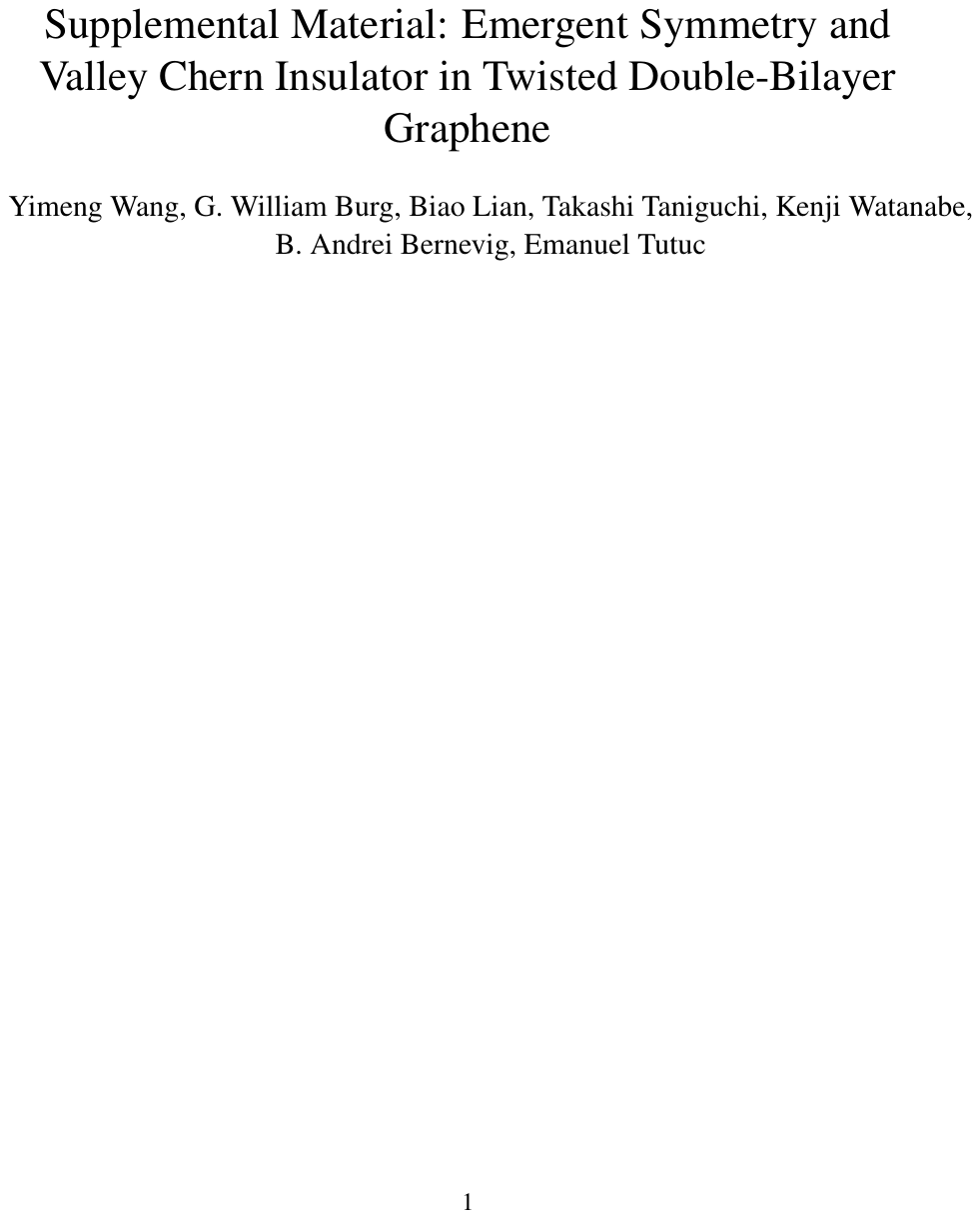}
{\color{white} .}
\newpage
\includepdf[pages={2}]{SM.pdf}
{\color{white} .}
\newpage
\includepdf[pages={3}]{SM.pdf}
{\color{white} .}
\newpage
\includepdf[pages={4}]{SM.pdf}
{\color{white} .}
\newpage
\includepdf[pages={5}]{SM.pdf}
{\color{white} .}
\newpage
\includepdf[pages={6}]{SM.pdf}
{\color{white} .}
\newpage
\includepdf[pages={7}]{SM.pdf}
{\color{white} .}
\newpage
\includepdf[pages={8}]{SM.pdf}
{\color{white} .}
\newpage
\includepdf[pages={9}]{SM.pdf}
{\color{white} .}
\newpage
\includepdf[pages={10}]{SM.pdf}
{\color{white} .}
\newpage
\includepdf[pages={11}]{SM.pdf}
{\color{white} .}
\newpage
\includepdf[pages={12}]{SM.pdf}
{\color{white} .}
\newpage
\includepdf[pages={13}]{SM.pdf}
{\color{white} .}
\newpage
\includepdf[pages={14}]{SM.pdf}
{\color{white} .}
\newpage
\includepdf[pages={15}]{SM.pdf}
{\color{white} .}
\newpage
\includepdf[pages={16}]{SM.pdf}
{\color{white} .}
\newpage
\includepdf[pages={17}]{SM.pdf}
{\color{white} .}
\newpage
\includepdf[pages={18}]{SM.pdf}
{\color{white} .}
\newpage
\includepdf[pages={19}]{SM.pdf}
{\color{white} .}
\newpage
\includepdf[pages={20}]{SM.pdf}
{\color{white} .}
\newpage
\includepdf[pages={21}]{SM.pdf}
{\color{white} .}
\newpage
\includepdf[pages={22}]{SM.pdf}
{\color{white} .}
\newpage
\includepdf[pages={23}]{SM.pdf}
{\color{white} .}
\newpage
\includepdf[pages={24}]{SM.pdf}
{\color{white} .}
\newpage
\includepdf[pages={25}]{SM.pdf}
{\color{white} .}

\end{document}